\documentclass[aps,pra,amsmath,amssymb,amsfonts,reprint,a4paper,superscriptaddress]{revtex4-1}
\pdfoutput=1 
\usepackage[T1]{fontenc} 

\usepackage{newtxtext,newtxmath}
\usepackage[utf8]{inputenx} 
\usepackage{graphicx}
\usepackage[dvipsnames]{xcolor}
\usepackage{soul} 
\usepackage[pdftex]{hyperref}
\usepackage{booktabs} 
\usepackage{multirow} 
\usepackage{lineno} 

\hypersetup{pdfauthor={Thomas Auzelle}, bookmarksnumbered=true, pdftitle={External control of GaN band bending using phosphonate self-assembled monolayers}, colorlinks, citecolor=blue, linkcolor=blue, urlcolor=blue}

\usepackage{gensymb}

\newcommand{\degC}{\,$^\circ$C}

\begin{document}
\title{External control of GaN band bending using phosphonate self-assembled monolayers}

\author{T.~Auzelle} 
\affiliation{Paul-Drude-Institut für Festkörperelektronik, Leibniz-Institut im Forschungsverbund Berlin e.\,V., Hausvogteiplatz 5--7, 10117 Berlin, Germany}
\email{auzelle@pdi-berlin.de}

\author{F.~Ullrich}
\altaffiliation{Materials Science Department, Technische Universit\"at Darmstadt, Otto-Berndt-Strasse 3, 64287 Darmstadt, Germany}
\affiliation{InnovationLab, Speyerer Str. 4, 69115 Heidelberg, Germany}

\author{S.~Hietzschold}
\altaffiliation{Institute for High-Frequency Technology, Technische Universit\"at Braunschweig, Schleinitzstrasse 22, 38106 Braunschweig, Germany}
\affiliation{InnovationLab, Speyerer Str. 4, 69115 Heidelberg, Germany}

\author{C. Sinito}
\altaffiliation{Present address: Attolight AG, EPFL Innovation Park, Bât.\ D, 1015 Lausanne, Switzerland}
\affiliation{Paul-Drude-Institut für Festkörperelektronik, Leibniz-Institut im Forschungsverbund Berlin e.\,V., Hausvogteiplatz 5--7, 10117 Berlin, Germany}

\author{S.~Brackmann}
\affiliation{InnovationLab, Speyerer Str. 4, 69115 Heidelberg, Germany}

\author{W.~Kowalsky}
\altaffiliation{Institute for High-Frequency Technology, Technische Universit\"at Braunschweig, Schleinitzstrasse 22, 38106 Braunschweig, Germany}
\altaffiliation{Kirchhoff Institute for Physics, Heidelberg University, Im Neuenheimer Feld 227, 69120 Heidelberg, Germany}
\affiliation{InnovationLab, Speyerer Str. 4, 69115 Heidelberg, Germany}


\author{E.~Mankel}
\altaffiliation{Materials Science Department, Technische Universit\"at Darmstadt, Otto-Berndt-Strasse 3, 64287 Darmstadt, Germany}
\affiliation{InnovationLab, Speyerer Str. 4, 69115 Heidelberg, Germany}

\author{O. Brandt}
\affiliation{Paul-Drude-Institut für Festkörperelektronik, Leibniz-Institut im Forschungsverbund Berlin e.\,V., Hausvogteiplatz 5--7, 10117 Berlin, Germany}

\author{R.~Lovrincic}
\altaffiliation{Institute for High-Frequency Technology, Technische Universit\"at Braunschweig, Schleinitzstrasse 22, 38106 Braunschweig, Germany}
\altaffiliation{Present address: trinamiX GmbH, Industriestraße 35, 67063 Ludwigshafen, Germany}
\affiliation{InnovationLab, Speyerer Str. 4, 69115 Heidelberg, Germany}

\author{S.~Fern\'andez-Garrido}
\altaffiliation{Grupo de Electr\'onica y Semiconductores, Dpto. F\'isica Aplicada, Universidad Aut\'onoma de Madrid, C/ Francisco Tom\'as y Valiente 7, 28049 Madrid, Spain}
\email{sergio.fernandezg@uam.es}
\affiliation{Paul-Drude-Institut für Festkörperelektronik, Leibniz-Institut im Forschungsverbund Berlin e.\,V., Hausvogteiplatz 5--7, 10117 Berlin, Germany}

\begin{abstract}
We report on the optoelectronic properties of GaN$(0001)$ and $(1\bar{1}00)$ surfaces after their functionalization with phosphonic acid derivatives. To analyze the possible correlation between the acid's electronegativity and the GaN surface band bending, two types of phosphonic acids, n-octylphosphonic acid (OPA) and 1H,1H,2H,2H-perfluorooctanephosphonic acid (PFOPA), are grafted on oxidized GaN$(0001)$ and GaN$(1\bar{1}00)$ layers as well as on GaN nanowires. The resulting hybrid inorganic/organic heterostructures are investigated by X-ray photoemission and photoluminescence spectroscopy. The GaN work function is changed significantly by the grafting of phosphonic acids, evidencing the formation of dense self-assembled monolayers. Regardless of the GaN surface orientation, both types of phosphonic acids significantly impact the GaN surface band bending. A dependence on the acids' electronegativity is, however, only observed for the oxidized GaN$(1\bar{1}00)$ surface, indicating a relatively low density of surface states and a favorable band alignment between the surface oxide and acids' electronic states. Regarding the optical properties, the covalent bonding of PFOPA and OPA on oxidized GaN layers and nanowires significantly affect their external quantum efficiency, especially in the nanowire case due to the large surface-to-volume ratio. The variation in the external quantum efficiency is related to the modication of both the internal electric fields and surface states. These results demonstrate the potential of phosphonate chemistry for the surface functionalization of GaN, which could be exploited for selective sensing applications.
\end{abstract}

\maketitle

Adsorbates on the surface of an inorganic semiconductor can perturb the equilibrium of charges established between bulk and surface states \cite{Darling1991}. Such a perturbation modifies the electrostatic potential landscape within the semiconductor, likely with sizeable consequences on carrier transport and recombination. In the case of GaN, the surface band bending has been reported to strongly depend on the adsorption of H$^+$ and OH$^-$ radicals \cite{Steinhoff2003a}, a phenomenom used to develop highly sensitive nitride-based pH sensors \cite{Wallys2012,Teubert2013}. Interestingly, even the sole physisorption of molecular acceptors \cite{Schultz2016} or a direct contact with electrolytes of different ionic strength \cite{Podolska2010} is sufficient to alter the internal GaN electrostatic potential. 
This long range electrochemical perturbation, propagating well below the GaN surface, is exploited in nitride-based bio and chemo-sensors. The probing area generally consists in the $(0001)$ facet of an ungated high-electron-mobility transistor (HEMT). In such a device, the adsorption events modify the charge density of the 2D electron gas (2DEG), and thus the transistor conductivity \cite{Anvari2018}. To increase the device surface-to-volume ratio and, thereby, boost device sensitivity, nanowire-based sensors are also under investigation \cite{Li2017}. In this scheme, the adsorption events on the nanowire sidewalls may translate into a modification of the nanowire luminescence, electrical conductivity and/or voltametric signature.
To further enable a selective sensing, an organic coating only binding to targeted analytes can be deposited on the GaN surface \cite{Kirste2015}. A low density of interface states, at the hybrid inorganic/organic interface, is then required to preserve the surface band bending sensitivity toward immobilized adsorbates \cite{Darling1991}. 

In this context, well-defined hybrid interfaces should be preferred, as those provided by self-assembled monolayers (SAMs). A SAM corresponds to a densely packed single layer of molecules, all covalently bonded to the surface through an anchor group. Each molecule additionally features a functional group, independent from the anchor group and adequately chosen for tailoring the surface properties. SAMs on metals and semiconductors (organic and inorganic) are widely reported and recurrently used as immobilizers as well as to tune the surface work function and the surface wettability \cite{Schlesinger2015,Casalini2017}. Because the oxidation of bare GaN surfaces is thermodynamically favored under ambient exposure \cite{Bermudez2017}, covalent binding of molecules on unoxidized GaN surfaces requires to work in ultra-high vacuum conditions. A more practical alternative is to deposit the SAM directly on the oxidized surface, thus enabling operation in air. In this regard, it is known that phosphonic acid can be attached to the GaN native oxide with various consequences for the optoelectronic properties of the underlying GaN layer, notably a modulation of its photoluminescence intensity and a shift of the core levels \cite{Wilkins2014,Wilkins2015a,Wilkins2015b}. After hydroxylation of the surface, phosphonic acids are reported to form dense and robust SAMs on the oxidized GaN$(0001)$ surface \cite{Ito2008,Kim2008}. Nevertheless, once grafted on the $(0001)$ gate of a HEMT, the functional group of these phosphonate SAMs only marginally affects the 2DEG density. This unsuitable behaviour is attributed to the presence of a large density of states at the GaN/GaO$_x$/SAMs interfaces \cite{Simpkins2010}. To date, a comparable analysis on the GaN$\{1\bar{1}00\}$ facets has not been accomplished. Such facets are, however, of high practical interest as they constitute the sidewalls \cite{Chen2001,Largeau2008} of most types of group-III nitride nanowires envisioned for bio-sensing applications \cite{Li2017}. 

Here, we analyze the optoelectronic properties of GaO$_x$/GaN$(0001)$ and $(1\bar{1}00)$ surfaces after their functionalization by grafting phosphonic acids. In particular, our work seeks to establish a correlation between the GaN surface band bending and the electronegativity of the attached acid, as theoretically proposed and experimentally demonstrated in the case of ZnO \cite{McNeill2016,Hofmann2017}, a material with various similarities to GaN with regard to crystal structure and related properties. To this end, we work with n-octylphosphonic acid (OPA) and 1H,1H,2H,2H-perfluorooctanephosphonic acid (PFOPA). These two different types of acids are characterized by a low and high electronegativity, respectively. Their self-assembly into a monolayer on oxidized GaN is confirmed by examining the related modification of the work function. In the GaO$_x$/GaN$(1\bar{1}00)$ case, we further evidence a substantial dependence of the surface band bending on the type of phosphanate SAM, in correlation with its electronegativity. Additionally, we evaluate the modification of the GaN external quantum efficiency (EQE) after grafting phosphonic acids. In this respect, beside GaN layers, we also examine GaN nanowires, for which surface related effects are typically enhanced. The measured EQE modifications are discussed in terms of surface band bending and surface states. The main conclusions derived from this study evidence the potential of the phosphonate chemistry to functionalize GaN nanowires into selective chemo-sensors. 

\section{EXPERIMENTAL DETAILS}

\begin{figure}
\includegraphics[width = \linewidth]{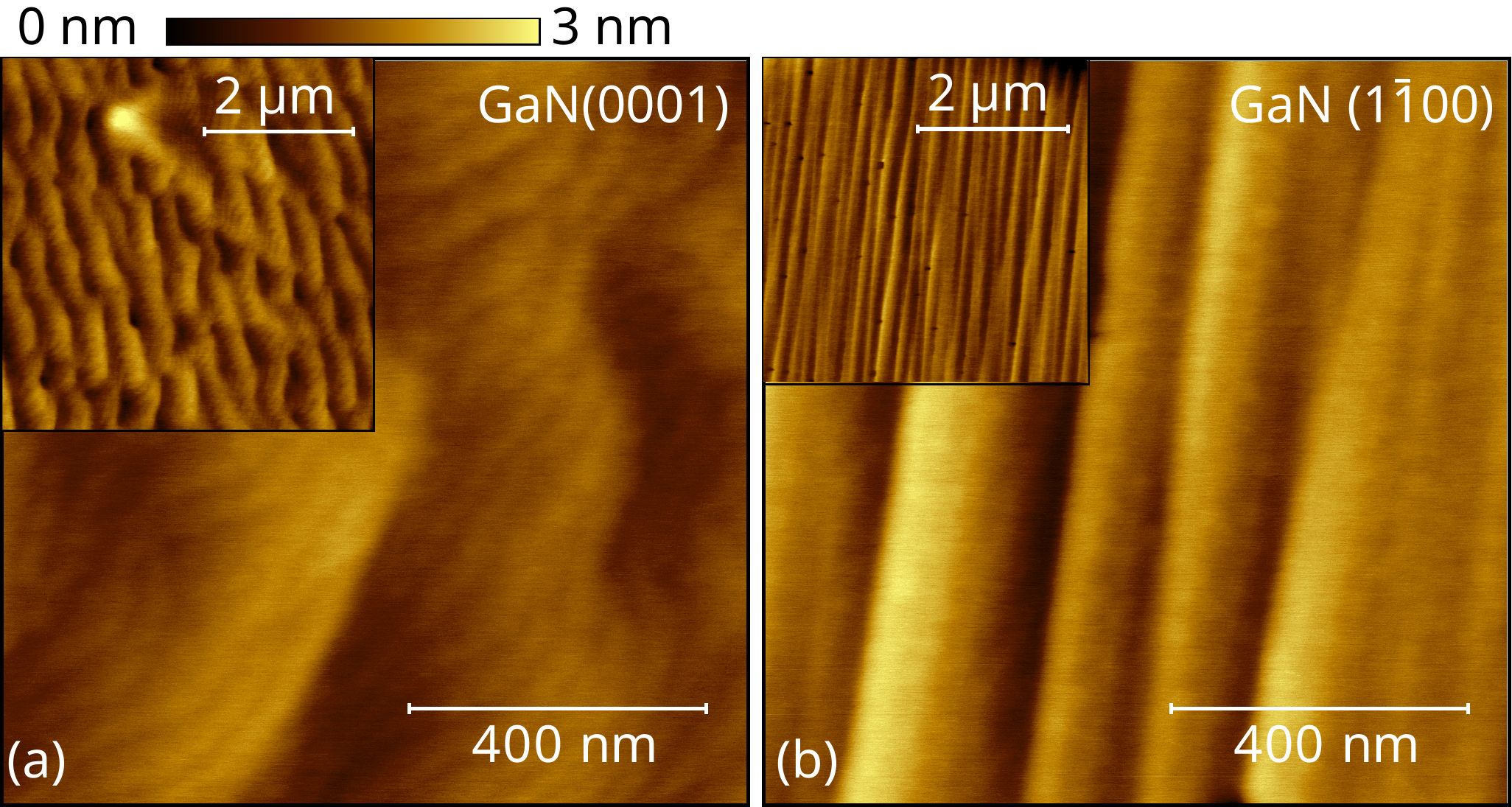}%
\caption{Representative atomic force micrographs of the PA-MBE regrown (a) GaN$(0001)$ and (b) GaN$(1\bar{1}00)$ surfaces under study. 
\label{fig:samples_layer}}
\end{figure}

The GaN$(0001)$ and $(1\bar{1}00)$ surfaces studied in this work are obtained from free-standing substrates purchased from Suzhou Nanowin Science and Technology. These substrates feature a low threading dislocation density ($< 5 \times 10^6$\,cm$^{-2}$) and a root mean square roughness below $0.2$\,nm. The surfaces are prepared by the vendor using chemo-mechanical polishing, which does not necessarily produce well-defined surfaces with atomic steps. Thus, to ensure reproducible surfaces, we grow an $\approx 1$\,\textmu m thick GaN layer by plasma-assisted molecular beam epitaxy (PA-MBE) on the as-received free-standing substrates. To preserve the substrate smoothness, the growth is performed at $\approx 700$\degC{} under intermediate Ga-rich conditions \cite{Heying2000,Shao2013,Lim2017}. Representative atomic force micrographs (AFM) of the MBE regrown GaN$(0001)$ and $(1\bar{1}00)$ layers are shown in Fig.\,\ref{fig:samples_layer}. Regardless of the surface orientation, the micrographs confirm the smoothness of the surfaces and the presence of atomic steps. The layers are unintentionally doped but exhibit a concentration of donors minus acceptors in the upper $10^{16}$\,cm$^{-3}$, as measured by capacitance voltage profiling using a two-point Hg probe. This residual \emph{n}-type doping is caused by the unintentional incorporation of O in PA-MBE.

\begin{figure}
\includegraphics[width = \linewidth]{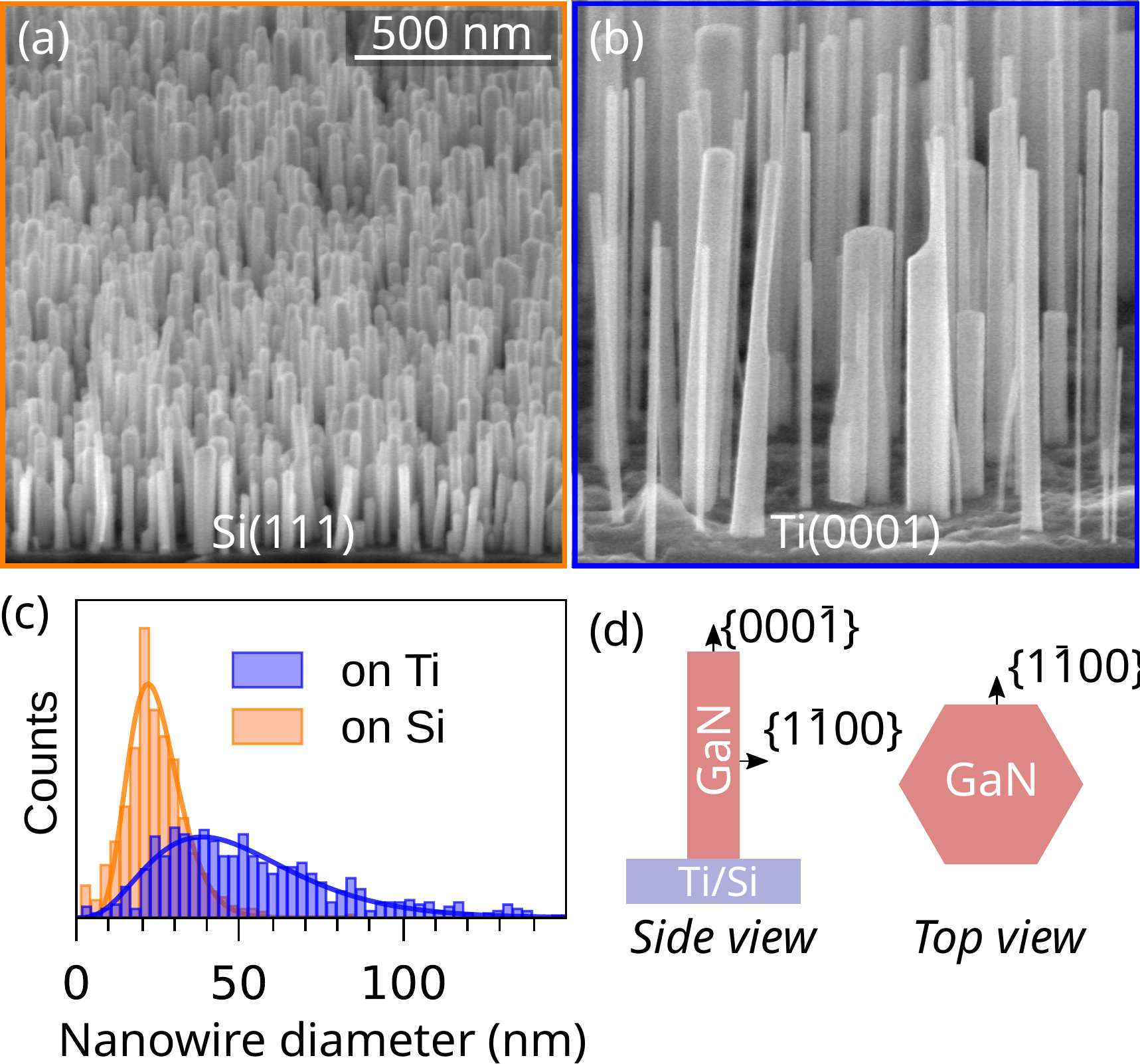}%
\caption{Bird's eye view scanning electron micrographs of the GaN nanowire ensembles grown by PA-MBE (a) on Si and (b) on Ti/Al$_2$O$_3$(0001). The magnification is the same in both micrographs. (c) Diameter distributions, as derived from the analysis of top-view scanning electron micrographs, for the two GaN nanowire ensembles under scrutiny. The solid lines are fits of a shifted gamma distribution to the histograms. (d) Schematic representation of the nanowires' main crystalline directions.
\label{fig:samples_NWs}}
\end{figure}

In addition to the layers, we prepare two GaN nanowire ensembles by PA-MBE using self-assembly as growth approach \cite{Consonni2013review}. The first ensemble is grown on deoxidized Si(111) at $805$\degC{} with a Ga flux of $3.6\times 10^{14}$\,cm$^{-2}$s$^{-1}$ and a N flux of $7.5\times10^{14}$\,cm$^{-2}$s$^{-1}$. To prevent the coalescence of nanowires nucleated in the early stage of growth, \cite{Kaganer2016a} the growth is interrupted before the full completion of the so-called ``nucleation phase''  \cite{Garrido2015}. The second nanowire ensemble is prepared on a Ti layer sputtered on Al$_{2}$O$_{3}$(0001). This ensemble is grown at $\approx 780$\degC{} with Ga and N fluxes of $4.8\times 10^{14}$ and $1.2\times10^{15}$\,cm$^{-2}$s$^{-1}$, respectively. As advocated in Ref.~\citenum{Calabrese2019}, prior to nanowire growth, the Ti surface is exposed to active N at $100$\degC{} to mitigate Ti, Al$_2$O$_3$ and Ga intermixing. On such a substrate, nanowires nucleate with a relatively low density, so that coalescence by bundling with neighboring nanowires is reduced \cite{vanTreeck2018}. In general, nanowire coalescence is undesirable since it generates extended defects within the nanowires \cite{Grossklaus2013,Kaganer2016b}, alters their prismatic shape \cite{Brandt2014} and induces strain \cite{Jenichen2011,Garrido2014,Auzelle2016}. Each of these contributions will decrease the uniformity of the nanowire ensembles and thus scatter the response of individual nanowires to the surface modifications investigated here. Scanning electron micrographs of the two different nanowire ensembles are respectively shown in Figs.\,\ref{fig:samples_NWs}(a) and \ref{fig:samples_NWs}(b). The nanowires prepared on Si have an average length of $200$\,nm and a density of $1.3 \times 10^{10}$ \,cm$^{-2}$. On Ti, the nanowires have an average length of 1\,\textmu m and a density of $4 \times 10^9$\,cm$^{-2}$. A shifted gamma distribution accurately fits the nanowire diameter distributions depicted in Fig.\,\ref{fig:samples_NWs}(c). The mean diameter derived from the fit for the nanowires grown on Si (Ti) is $23$\,nm ($47$\,nm). As schematically shown in Fig.\,\ref{fig:samples_NWs}(d), such self-assembled nanowires generally feature $\{1\bar{1}00\}$ sidewalls \cite{Largeau2008} and a $(000\bar{1})$ top-facet \cite{ZunigaPerez2016,Calabrese2020a}.

\begin{figure}
\includegraphics[width = \linewidth]{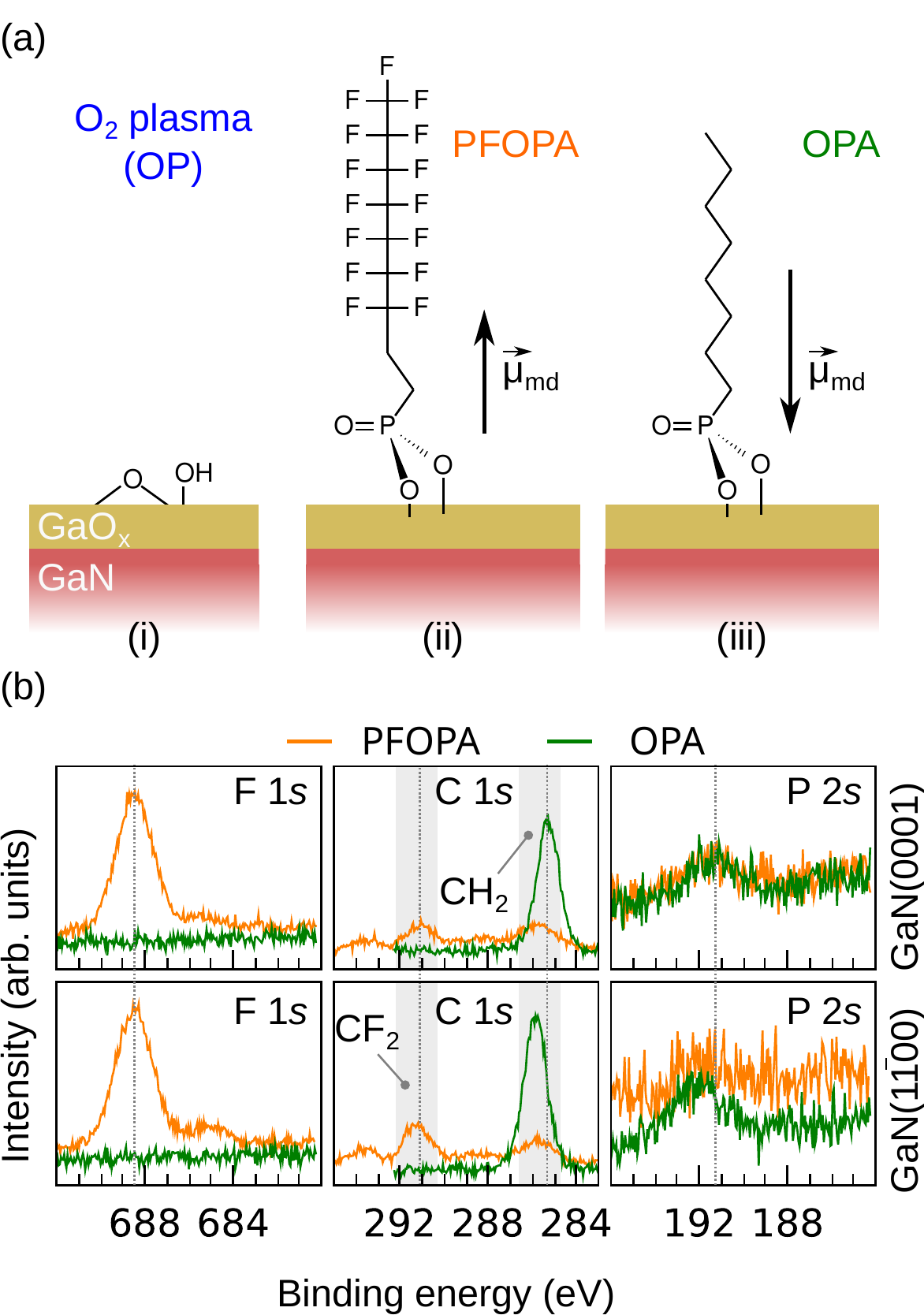}%
\caption{(a) Exemplary view of the GaN surface after (i) the OP treatment and once (ii) PFOPA or (iii) OPA are grafted on the oxidized GaN surface in a bidentate configuration. $\vec{\mu}_{md}$ indicates the direction of the molecular dipole relative to the anchor-group. (b) F\,$1s$, C\,$1s$ and P\,$2s$ core level intensity for GaN$(0001)$ and GaN$(1\bar{1}00)$ surfaces after OP treatment and upon grafting of PFOPA or OPA.
\label{fig:molecules}}
\end{figure}

Prior to the SAM deposition, both the GaN layers and the nanowire ensembles are cleaned and oxidized using the following subsequent steps: O$_2$ plasma (OP) exposure, HCl etching, and a second OP exposure. In detail, the OP exposure step is $5$\,min long and performed using a O$_2$ pressure of $0.3$\,mbar and a forward power of $900$\,W. For the HCl etching step, the samples are immersed for $10$\,min in a $32$\% HCl solution. Before the characterization of the samples by photoluminescence spectroscopy (Section\,\ref{sec:PL}), we use, however, a different OP system. In this case, the OP exposure is $10$\,min long and carried out using an O$_2$ pressure of $0.5$\,mbar and a forward power of $100$\,W. Besides the elimination of pollutants, the cleaning process also initializes the surface into a well-defined reference state, which is characterized by the formation of a $\leq1$\,nm thick GaO$_x$ capping layer \cite{Auzelle2019}. Partial hydroxylation of the surface is assumed, which favors the anchoring of phosphonic acids \cite{Gardner1995,Perkins2009}. 
PFOPA ($\geq95$\%, SigmaAldrich) and OPA molecules ($99$\%, AlfaAeasar) [sketched in Fig.\,\ref{fig:molecules}(a)] as well as anhydrous ethanol ($\geq 99.8$\%, VWR) are used as received. The phosponic acids are dissolved in anhydrous ethanol with a concentration of $2$\,mmol/L and stirred for at least $3$h at $30$\degC{}. The GaN samples are then immersed in the solution for $3$h at $50$\degC{} to allow the physisorption of the molecules. Subsequently, the samples are dried in air on a hot plate at $140$\degC{} to foster the chemisorption of the physisorbed phosphonic acids. At last, the samples are thoroughly rinsed in ethanol and dried with a N$_2$ gun to remove non-chemisorbed molecules. 

The electronic properties of the GaN surfaces are probed by X-ray photoelectron spectroscopy (XPS) in a PHI Versa Probe II equipped with a monochromatized Al K$_{\alpha}$ anode as X-ray source ($1486.6$\,eV). The base pressure of the measuring chamber is in the $10^{-9}$\,mbar range. The kinetic energy of the photo-emitted electrons is measured by a concentric hemispherical analyzer. The angle between the analyzer and the X-ray source is $54.7^\circ$ and the takeoff angle $90^\circ$. 
All the spectra are referenced in binding energy with respect to the Fermi edge of Ag. The secondary electron edges are calibrated according to the work function of an in situ sputter-cleaned Ag foil. The work function of the Ag foil is determined by ultraviolet photoelectron spectroscopy. XPS fits are performed using Voigt line profiles after subtraction of a Shirley background \cite{Shirley1972}.

The optical properties of the GaN samples are characterized by continuous-wave photoluminescence spectroscopy (cw-PL) using the $325$~nm line of a He-Cd laser. A lens with a numerical aperture of $0.3$ focuses the laser beam onto the sample surface with a spot size of $\approx 50$\,\textmu m. When measuring nanowire ensembles, $1$--$3 \times 10^5$ nanowires are then probed at the same time. The PL is collected in backscatter geometry, dispersed by a $80$~cm monochromator, and detected by a charge-coupled-device (CCD) detector cooled down with liquid nitrogen. The samples are mounted with silverpaint on a cold-finger cryostat for measurement at cryogenic temperatures. All measurements are performed in a low $10^{-6}$\,mbar vacuum. The excitation density is attenuated using neutral density filters.

\section{RESULTS and DISCUSSION}
\subsection{Self-assembled monolayer formation}

The tendency of PFOPA and OPA to self-assemble into a monolayer on oxidized GaN$(0001)$ and $(1\bar{1}00)$ surfaces is examined by XPS. Figure\,\ref{fig:molecules}(b) shows the C\,$1s$, P\,$2s$ and F\,$1s$ core levels after grafting of PFOPA or OPA on the free-standing GaN layers. The presence of the phosphonic acids is confirmed by the weak P\,$2s$ signal at $192$\,eV and the stronger C\,$1s$ signal, attributed to CH$_2$ \cite{Angel2014}, at $286$--$287$\,eV. In the particular case of PFOPA, a strong F\,$1s$ signal is additionally observed at $688.5$\,eV together with a second contribution to the C\,$1s$ signal at $291$--$293$\,eV. We attribute the latter signal to CF$_2$ \cite{Angel2014}. Remarkably, the core levels of OPA on GaN$(1\bar{1}00)$ have a $\approx 1$\,eV higher binding energy compared to those of OPA on GaN$(0001)$. This result indicates different degrees of charge transfer between the chemisorbed OPA molecules and the GaO$_x$ layer created on the $(1\bar{1}00)$ and $(0001)$ facets. Hence, OPA has a higher oxidation state (less electrons) on GaN$(1\bar{1}00)$ than on GaN$(0001)$.

\begin{figure}
\includegraphics[width = \linewidth]{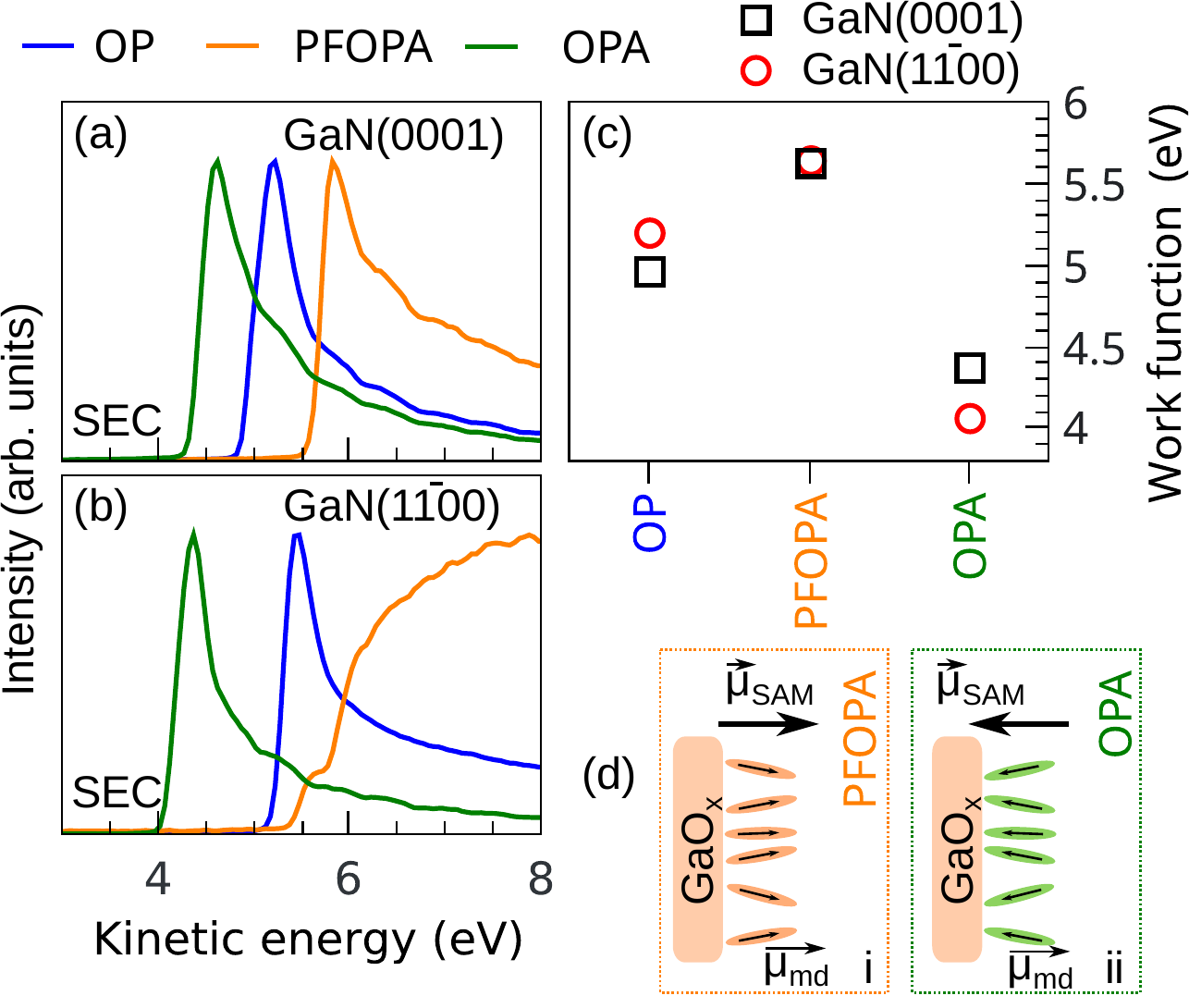}%
\caption{Secondary electron cutoff (SEC) for GaN$(0001)$ (a) and GaN$(1\bar{1}00)$ (b) after OP treatment and either grafted with PFOPA or OPA. (c) Work functions extracted from the SECs. (d) Schematic of the surface dipoles respectively built by (i) PFOPA and (ii) OPA.
\label{fig:WF}}
\end{figure}

We also derive by XPS the work function of the GaN layers grafted with OPA or PFOPA from the electron kinetic energy at the secondary electron cutoff (SEC). As shown in Fig.\,\ref{fig:WF}, for both GaN orientations, PFOPA grafting increases the work function to $5.6$\,eV, whereas OPA grafting lowers it to $4.1$--$4.4$\,eV. We attribute these modifications to a macroscopic surface dipole ($\vec{\mu}_{SAM}$) build-up by the phosphonic acids standing-up on the GaO$_x$ surface \cite{Hofmann2010} [see sketch in Fig.\,\ref{fig:WF}(d)]. The opposite polarity of the surface dipoles respectively created by PFOPA and OPA relates to the opposite orientation of the molecular dipoles ($\vec{\mu}_{md}$) hosted by the grafted molecules. In view of the large amplitude of the surface dipole ($\approx 0.5$\,eV), we conclude that both types of phosphonic acids self-assemble into a dense monolayer. Only in the case of GaN$(1\bar{1}00)$ coated with PFOPA, the broad SEC suggests the presence of surface inhomogeneities, possibly due to the formation of an incomplete SAM. Hence, as for the commonly used GaN$(0001)$ orientation \cite{Ito2008,Kim2008,Simpkins2010}, phosphonate acids also self-assemble into a monolayer on the GaN$(1\bar{1}00)$ surface. This process is of high practical interest for achieving surface functionalization of GaN nanowires.  

\subsection{Tailoring of surface band bending}
\begin{figure*}
\includegraphics[width = \linewidth]{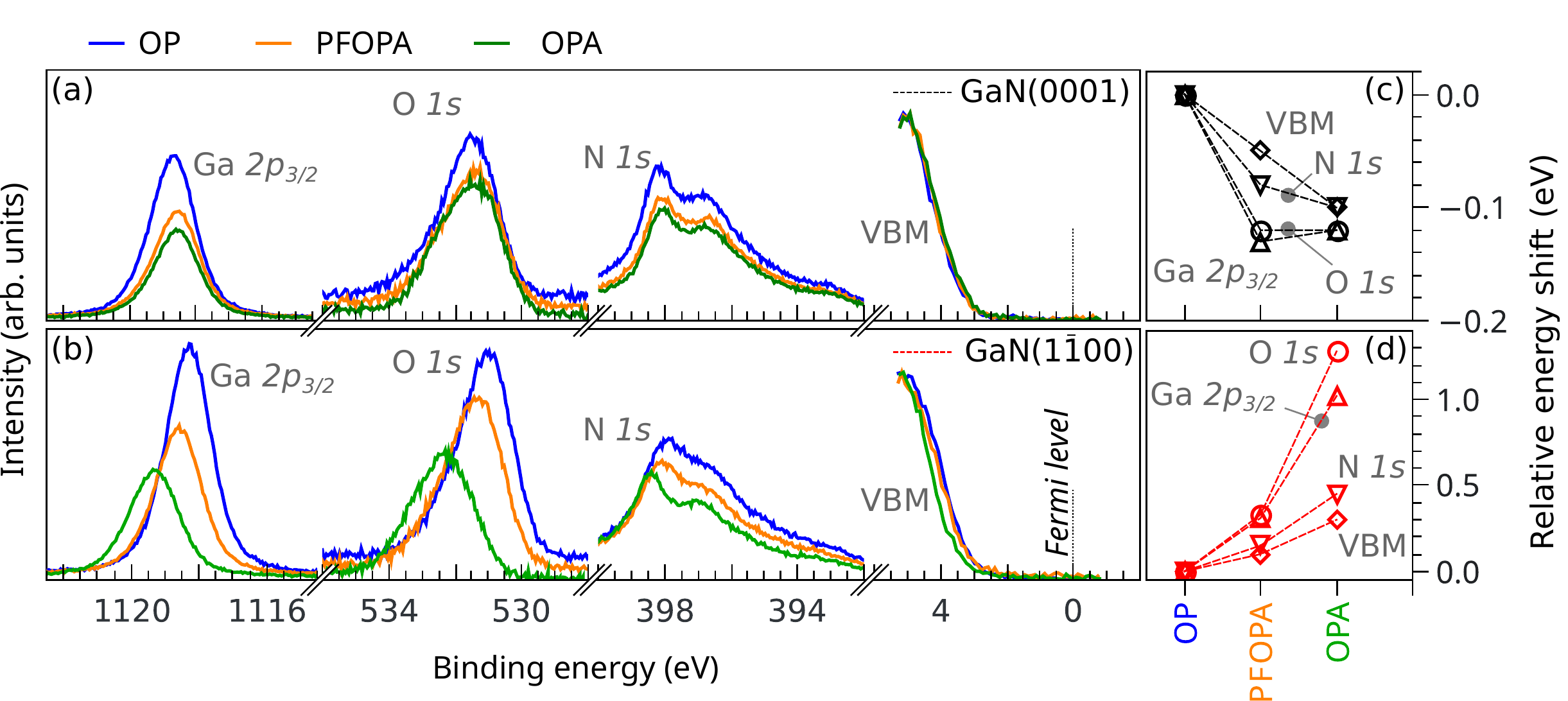}%
\caption{(a)--(b) Binding energies of the Ga\,$2p_{3/2}$, O\,$1s$ and N\,$1s$ core levels and of the valence band maximum (VBM) measured by XPS after OP and upon grafting of PFOPA or OPA. Only the VBM onset is normalized. (c)--(d) Shift of the core levels and VBM relative to the OP treated surface. (a)--(c) refers to the GaN$(0001)$ and (b)--(d) to GaN$(1\bar{1}00)$.
\label{fig:BB}}
\end{figure*}

The presence of a SAM can affect the surface band bending, since it will modify the equilibrium of charges existing between GaN donor states and GaO$_x$ surface states \cite{Darling1991,Hofmann2017,Anvari2018}. To analyze the impact of the phosphonic acid SAMs on the surface band bending, we assess by XPS the energy difference between the Fermi level and the valence band maximum (VBM) as well as between the Fermi level and the Ga, O and N core levels. These energy differences are hereafter referred to as the binding energy of the VBM and the core levels, respectively. These measurements are carried out for the GaN$(0001)$ and $(1\bar{1}00)$ layers after OP exposure and upon grafting with PFOPA or OPA. The results are shown in Figs.\,\ref{fig:BB}(a) and \ref{fig:BB}(b). In the following, the SAM-induced modification of the GaN/GaO$_x$ core levels is systematically discussed with respect to the values of the OP treated surface, which are thus taken as reference. 

For both GaN surface orientations, in the presence of PFOPA or OPA, the core levels show a reduced intensity and the O\,$1s$ signal is asymmetrically broadened toward larger binding energies. Since the inelastic mean free path of the core electrons is $1\text{--}2$\,nm \cite{Krawczyk2004}, the inelastic scattering undergone by these electrons when escaping through the phosphonate SAM can explain the reduced core level intensity. With respect to the O\,$1s$ signal, its asymmetric broadening is attributed to the formation of O--P bonds at the surface, as reported elesewhere  \cite{Ito2008}.
More interestingly, the grafting of the phosphonic acids induces a shift of the core levels and VBM binding energies. The recorded shifts are plotted in Figs.\,\ref{fig:BB}(c) and \ref{fig:BB}(d). Each core level experiences a different displacement, a result that differs from previous studies on ZnO \cite{McNeill2016}. These shifts results from the combination of two different phenomena: (i) a modification of the GaN surface band bending, and (ii) a change in the oxidation state of the atoms forming the GaO$_x$ layer. 


For undoped or lightly doped material, the surface band bending extends over a length scale $\gg10$\,nm, a value that largely exceeds the probing depth of XPS ($1\text{--}2$\,nm). A modification of the surface band bending would then rigidly shift the GaN/GaO$_x$ core levels and the VBM binding energies. In contrast, the covalent bonding of the phosphonic acids at the GaO$_x$ surface can alter the oxidation state of the Ga and O atoms differently. Consequently, as the result of such an effect, different shifts are expected for the core levels and the VBM binding energies. 

To isolate the contribution of the GaN surface band bending to the core level shifts shown in Figs.\,\ref{fig:BB}(c) and \ref{fig:BB}(d), we first make two assumptions: (1) The GaO$_x$ layer is free of N atoms. This assumption is supported by experimental studies of deep plasma oxidation of GaN \cite{Pal2003,Lee2008} as well as by the theoretical modelling of an oxidized GaN surface \cite{Miao2010}. (2) The grafting of the phosphonic acids only impacts the oxidation state of the Ga and O atoms located in the GaO$_x$ layer. According to these two assumptions, the relative changes in the amplitude of the surface band bending scale one to one with the shifts of the N\,$1s$ core level energy. Last, we use the values reported in Ref.~\citenum{Auzelle2019} to obtain an absolute estimate of the surface band bending after the initial OP treatment. In this way, we are able to deduce the amplitude of the GaN surface band bending after the deposition of the different SAMs. The values obtained are compiled in Table\,\ref{table:BB}. 

\begin{table}
    \caption{Amplitudes of the upward GaN band bending ($\Phi_0$) extracted from the XPS measurements performed on the GaN$(0001)$ and $(1\bar{1}00)$ layers after OP as well as upon grafting of PFOPA or OPA. As described in the text, the width of the exciton dead layer ($w_{\text{edl}}$) is estimated for each cases.}
    \label{table:BB}
    \small{
    \begin{tabular}{lc||ccc}
        \multicolumn{2}{l}{Surface treatment} & \textbf{OP}       & \textbf{PFOPA}    & \textbf{OPA} \\\midrule\midrule
        \multicolumn{5}{c}{GaN$(1\bar{1}00)$ layer}\\
        $\Phi_0$                & $\pm 0.05$\,eV     ~~~& ~~~$0.50$     ~~~ &~~~$0.30$   ~~~&~~~$0.05$~~~  \\
        $w_{\text{edl}}$                & nm                     & ~~~$51$     ~~~ &~~~$38$   ~~~&~~~$12$~~~  \\      
        \multicolumn{5}{c}{}\\\hline
        \multicolumn{5}{c}{GaN$(0001)$ layer}\\
        $\Phi_0$                &$\pm 0.05$\,eV         & $0.10$            &$0.20$         &$0.20$  \\
        $w_{\text{edl}}$                & nm                     & ~~~$20$     ~~~ &~~~$31$   ~~~&~~~$31$~~~  \\       
    \end{tabular}
    }
\end{table}
For GaN$(0001)$, the deposition of either PFOPA or OPA SAMs induces moderate core level shifts ($<0.2$\,eV) and increases the surface band bending by $\approx0.1$\,eV. The surface band bending is thus not very sensitive to the electronegativity of the SAM functional group for GaN$(0001)$. In contrast, the grafting of phosphonic acids on GaN$(1\bar{1}00)$ largely affects both the core levels and the VBM. Specifically, after grafting of PFOPA, the GaN surface band bending is reduced by $0.2$\,eV and the oxidation states of Ga and O atoms exhibit a slight increase. Relative to PFOPA, the coating with OPA further decreases the GaN surface band bending by $0.25$\,eV, and increases the oxidation states of Ga and O atoms. These results evidence a strong impact of the SAM electronegativity on the inner GaN electrostatic potential landscape in the case of the $(1\bar{1}00)$ surface. 

A similarly weak sensitivity of the GaN$(0001)$ surface band bending toward the electronegativity of grafted phosphonic acids has been reported for oxidized GaN$(0001)$ \cite{Ito2008} and (Al,Ga)N$(0001)$ layers \cite{Simpkins2010}. This lack of sensitivity was explained in terms of a Fermi level pinning caused by a high density of states at the nitride surface. Nevertheless, beside the density of surface states, their specific energy with respect to the electronic states of the organic adsorbate will also control the polarity and amplitude of the charge transfer process occuring at the inorganic/organic interface \cite{Hofmann2017}. In particular, electron injection from OPA to the acceptor states of the surface GaO$_x$ can only occur if the latter have a lower energy compared to the highest occupied molecular orbital (HOMO) of OPA. Due to the strong anisotropy of the GaN crystal structure, such a band alignment can be entirely different for the GaN$(0001)$ and $(1\bar{1}00)$ orientations. The high sensitivity of the band bending of the oxidized GaN$(1\bar{1}00)$ layer to phosphonic acid grafting thus evidences a low density of surface states as well as a favorable band alignment for charge transfer at the inorganic/organic interface. Interestingly, the observed correlation of the band bending amplitude with the electronegativity of the phosphonic acid provides a convenient means for tailoring the electronic properties of the $(1\bar{1}00)$ surface. This result is particularly relevant for GaN nanowires because their electrical and optical properties are strongly influenced by the chemical properties of their $\{1\bar{1}00\}$ sidewall facets.  \cite{Calarco2005,Pfuller2010a,Gurwitz2011,Lefebvre2012,Wallys2012,BengoecheaEncabo2016,Maier2017,Hetzl2018} 

\subsection{Enhancement of the external quantum efficiency}\label{sec:PL}
The modification of the surface electronic properties caused by the grafting of phosphonic acids should also impact the EQE of GaN. In particular, the decrease of internal electric fields \cite{Blossey1970,Shokhovets2003,Winzer2006,Nelson2001,Camacho2003,Pedros2007} and the passivation of surface states \cite{Wilkins2015a,Wang2017,Latzel2017,Varadhan2017,Jafar2016,Zhao2015,Wong2018} could both enhance the radiative recombination of excitons. This issue is further evaluated here by examining the PL intensity of the coated GaN samples. In addition to the GaN$(0001)$ and GaN$(1\bar{1}00)$ layers, we also examine the GaN nanowire ensembles to benefit from their higher sensitivity to surface effects \cite{Pfuller2010a}. Measurements are carried out in vacuum ($\approx 10^{-6}$\,mbar) at $10$ and $300$\,K. All the samples are cut in three pieces. One of the pieces is treated with OP, and the other two are grafted with PFOPA or OPA, respectively. To compare the PL intensities, pieces from the same samples are loaded in the cryostat and measured side-by-side. 

\begin{figure*}
\includegraphics[width = .6\linewidth]{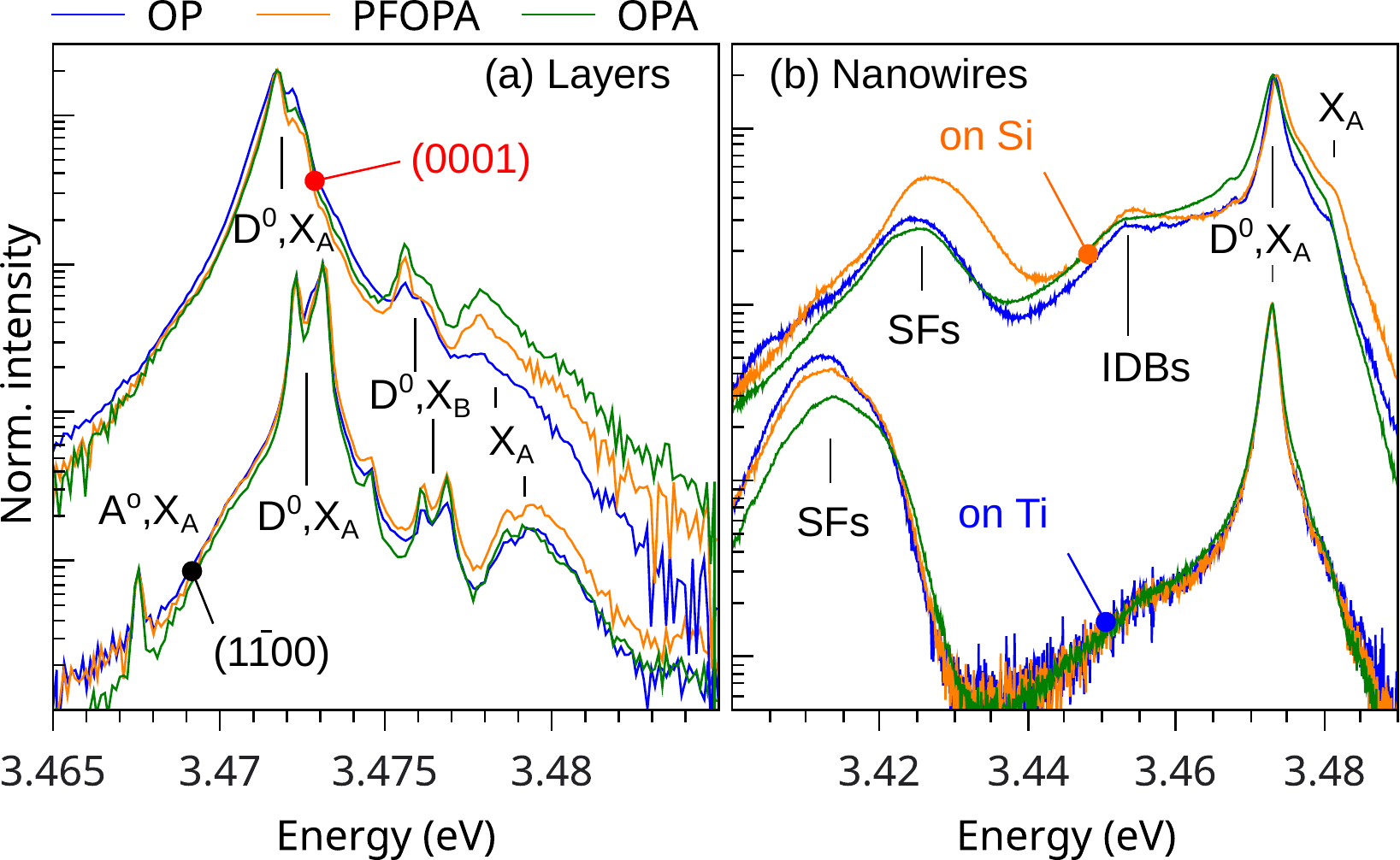}%
\caption{Normalized PL spectra of the (a) GaN layers and (b) nanowire ensembles taken at $10$\,K after OP as well as subsequent to the grafting of PFOPA or OPA. Spectra from different samples are vertically shifted for clarity. The different transitions are labeled according to Ref.~\citenum{Monemar2010}.
\label{fig:PLspectra}}
\end{figure*}

Both the treated layers and the nanowire ensembles exhibit  intense band-edge luminescence. Normalized spectra taken at $10$\,K are shown in Fig.\,\ref{fig:PLspectra}. In the case of the layers, the spectra feature narrow excitonic lines which are identified and labelled according to Ref.~\citenum{Monemar2010}. Excitonic lines are also observed in the nanowire ensembles, although broadened by the contribution of several surface-related effects \cite{Brandt2010,Corfdir2014b,Zettler2015,Calabrese2020a}. In particular, the donor bound exciton ($D^0$,$X_\text{A}$) has a linewidth of $1.5$\,meV ($4$\,meV) for the nanowire ensemble on Ti (Si). Transitions from excitons bound to stacking faults (SFs) \cite{Corfdir2014a} and inversion domain boundaries (IDBs) \cite{Auzelle2015,Pfuller2016} are also detected, as typically reported for this type of nanostructures. Interestingly, we do not observe any spectral signature related to the deposition of PFOPA or OPA. The differences visible in the spectra shown in Fig.\,\ref{fig:PLspectra} can be entirely explained in terms of sample inhomogeneities. A similar result is obtained at $300$\,K (not shown here). The absence of a shift in the exciton recombination energy is compatible with the altered internal electric field caused by the grafting of phosphonic acids \cite{Camacho2003,Pedros2007,Pfuller2010a}. As a matter of fact, for increasing electric fields, the radiative recombination of free excitons will quench ($F \geqslant 10$\,kV/cm) before a measurable redshift would occur ($F\geqslant 20$\,kV/cm \cite{Winzer2006}). The change of the internal electric fields should, however, affect the PL intensity.
 
\begin{figure*}
\includegraphics[width = .7\linewidth]{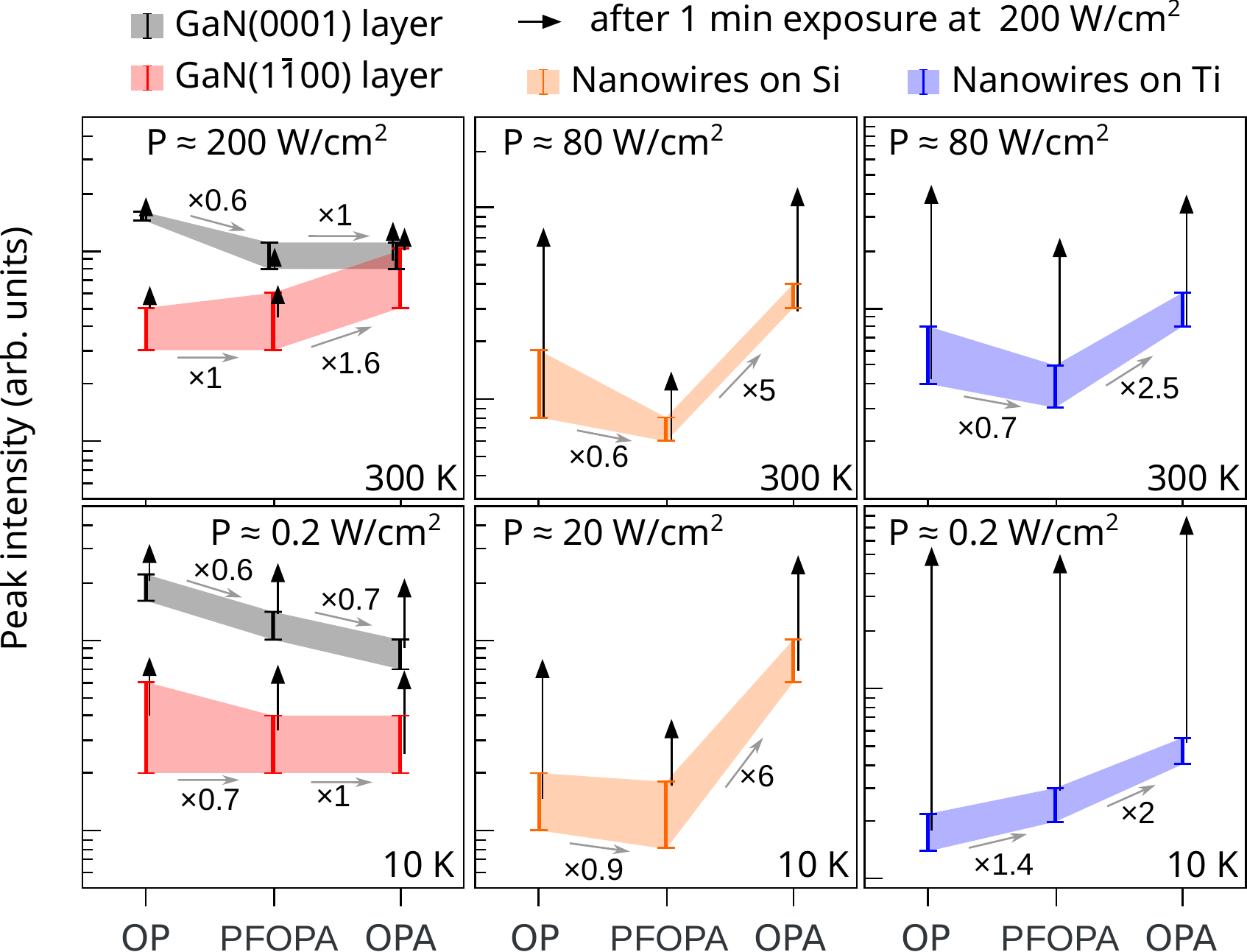}%
\caption{Peak PL intensity at 300 and  $10$\,K of the GaN layers and nanowires grown on Si and Ti after OP exposure as well as upon PFOPA or OPA grafting. The error bars highlighted by the colored areas represent the dispersion of the data along the sample. The excitation density used for the PL acquisition is given in each graph. The vertical black arrows indicate the intensity enhancement obtained after exposing the surface for a 1\,min to the maximum laser intensity of $200$\,Wcm$^{-2}$ prior to measuring the PL intensity. 
\label{fig:PL}}
\end{figure*}

The recorded PL peak intensities after the different treatments (OP, PFOPA and OPA) are shown in Fig.\,\ref{fig:PL} for the GaN layers and the nanowire ensembles. Under laser exposure, the luminescence intensity steadily increases for up to
several tens of minutes, as widely observed for as-grown GaN layers and nanowires \cite{Behn2000,Foussekis2009,Pfuller2010a,Hetzl2018}. To circumvent this complication, the values plotted in Fig.\,\ref{fig:PL} correspond to the first second of luminescence. An intensity range is given instead of an average intensity to reflect the dispersion of the values caused by sample inhomogeneities (within an area of $\approx10$\,mm$^2$). Besides these measurements, we also record the PL intensity in a location of the sample previously irradiated with the laser using an intensity of $200$\,W/cm$^2$ for $1$\,min. Such a process increases the PL intensity to the extent depicted by the black arrows shown in Fig.\,\ref{fig:PL}. For as-grown nanowires, this photo-induced enhancement is attributed to a chemical modification of the surface under laser irradiation, which reduces the surface band bending \cite{Pfuller2010a} and/or the areal density of nonradiative surface recombination centers \cite{Hetzl2018}. Nevertheless, different mechanisms can be expected for GaN surfaces with different chemical properties (for example, different oxide thickness and structure).

As shown in Fig.\,\ref{fig:PL}, the grafting of PFOPA and OPA induces moderate PL intensity changes for the layers and larger ones for the nanowire ensembles. This difference likely arises from the high nanowire surface-to-volume ratio. For the GaN$(1\bar{1}00)$ layer as well as for the nanowire ensembles, on average, OPA grafting enhances the PL intensity as compared to PFOPA. This increase amounts to up to a factor $6$ in the nanowire case. This PL enhancement correlates well with the expected reduction of the surface band bending (see Table\,\ref{table:BB}). Yet, PFOPA does not enhance the PL intensity in comparison to OP. According to the band bending reduction expected upon PFOPA coating (see Table\,\ref{table:BB}), this is an unexpected result that could be easily explained by taking into account the light absorbed by the organic overlayer. Last, the PL intensity enhancement observed after the intense laser exposure is remarkably unaffected by the presence of PFOPA and OPA. It thus likely corresponds to a photo-induced modification of the GaO$_x$ layer. 

The reduction of the surface band bending provided by PFOPA and OPA coatings shrinks the width of the space charge layer ($w_{\text{scl}}$). In the vicinity of the surface, this layer is associated with electric fields that may be high enough to field-ionize excitons ($F \gtrsim 10$\,kV/cm). We further estimate the width $w_{\text{edl}}$ of this ``exciton dead layer'' in the approximation of continuous charge distribution by solving the Poisson equation. In the planar case, $w_{\text{edl}} \in [0, w_{\text{scl}}]$ is given by:

\begin{equation}\label{eq:elec_layer}
    F(w_{\text{edl}}) = e \frac{N_D}{\varepsilon_r\varepsilon_0} \left( \frac{\Phi_0 \varepsilon_r\varepsilon_0}{e^2 N_D} - w_{\text{edl}} \right)^{\frac{1}{2}}= 10\text{ kV/cm}\\
\end{equation}
where $\Phi_0$ is the band bending amplitude, $N_D$ the donor concentration, and $\varepsilon_r \varepsilon_0$ the permittivity of GaN ($\varepsilon_r^{\parallel c} = 10.4$ and $\varepsilon_r^{\perp c} = 9.5$). The calculated values reported in Table\,\ref{table:BB} show that $w_{\text{edl}}$ has a similar order of magnitude than the absorption length of the laser light ($\approx 100$\,nm) and significantly decreases when reducing the surface band bending. Therefore, besides light absorption in the organic overlayer, we attribute the different PL intensities of the coated GaN layers to the variation of the width of the exciton dead layer.

For lightly doped nanowires, the amplitude of the surface band bending may additionally depend on their diameter. Indeed, below a critical diameter, full electronic depletion of the bulk donors by surface states occurs \cite{Calarco2005}. In such a case, the amplitude of the surface band bending depends on the diameter ($\Phi_0 \propto d^2$) and not anymore on the position of the Fermi level at the surface. This critical diameter $d_c$ can be written as:
\begin{equation}\label{eq:dcrit}
    d_{crit} = \sqrt{\frac{16 \varepsilon_r\varepsilon_0 \Phi_0 }{e^2N_D}}
\end{equation}
where $\Phi_0$ is the band bending amplitude when the nanowire is not fully depleted. Assuming a similar doping than for the GaN layers ($N_D = 8 \times 10^{16}$\,cm$^{-3}$), $d_c$ amounts to $230$,  $178$ and $72$\,nm after the OP, PFOPA and OPA treatments, respectively. Examining the diameter distributions shown in Fig.\,\ref{fig:samples_NWs}(c), we deduce that, even after grafting OPA, all the nanowires on Si and the vast majority of those grown on Ti are actually electronically fully depleted. For these nanowires, the organic coating would thus only marginally affect the surface band bending. A larger residual doping level is necessary to observe a surface dependence of the band bending. In particular, values in the $10^{18}$\,cm$^{-3}$ range would be required. These values are, however, incompatible with the narrow donor bound exciton transition observed in the nanowire cw-PL spectra ($\leq4$\,meV). We note that GaN nanowires doped with Si in the mid $10^{17}$\,cm$^{3}$ exhibit a donor-bound exciton transition with a linewidth of $7$\,meV \cite{Schlager2011}, and larger doping levels would further increase this value \cite{Schubert1997}. Hence, for thin nanowires ($d < 72$\,nm), such as for the ensemble prepared on Si, the increased PL intensity provided by OPA grafting cannot be due to a reduction of the width of the exciton dead layer.

Besides altering internal electric fields, the covalent bonding of phosphonic acids onto the GaO$_x$ surface certainly impacts the properties and nature of surface states, which may be a source for the nonradiative recombination of excitons. Yet, surface excitons only exists in nanowires with low surface electric fields ($F < 10$\,kV/cm), since they would otherwise be instantaneously field-ionized \cite{Corfdir2014b,Auzelle2019b}. Taking $N_D = 8 \times 10^{16}$\,cm$^{-3}$, surface excitons exist only in nanowires with a diameter below $26$\,nm. As seen in Fig.\,\ref{fig:samples_NWs}(c), these nanowires are actually the majority in the ensemble grown on Si, and represent a non-negligible fraction in the ensemble prepared on Ti. For these very thin nanowires, we thus expect a change of the PL intensity upon PFOPA or OPA coating due to a modification of the nonradiative surface recombination velocity. As discussed in the Supplementary Information, the surface defects involved in this surface recombination are not inherent to the as-grown GaN nanowires but primarily related to the OP treatment.

From this analysis, the change of the GaN nanowire EQE resulting from the grafting of PFOPA and OPA can be attributed to three different contributions: (i) for thick nanowires ($d > 72$\,nm), to a variation in the width of the exciton dead layer, (ii) for very thin nanowires ($d < 26$\,nm), to a modification of the surface recombination velocity, and (iii) regardless of the nanowire diameter, to the parasitic light absorption caused by the presence of the organic SAM. 
Consequently, nanowires with diameters between $26$ and $72$\,nm are unaffected by the first two contributions. Yet, we note that under large excitation power densities (higher than those used in this work) photo-screening of the band bending will occur \cite{Sezen2011,Sezen2014}, even for electronically depleted nanowires. Since the amplitude of the photo-screening is expected to depend on the nature of surface states \cite{Darling1991}, it will vary depending on the organic coating.   

\section{SUMMARY AND CONCLUSIONS}
We found that phosphonic acids equally bind to oxidized GaN$(0001)$ and $(1\bar{1}00)$ surfaces, resulting in both cases in the formation of dense SAMs. The binding is verified by the significant change of the GaN work function ($\pm\,0.5$\,eV) observed after the grafting of PFOPA or OPA. These two types of phosphonic acids furthermore affect the GaN surface band bending. For the oxidized GaN$(1\bar{1}00)$ surface, this effect scales with the electronegativity of the phosphonic acid's tail, a phenomenon that enables an external control of the GaN surface band bending. The oxidized GaN$(0001)$ surface is, in contrast, rather insensitive to the electronegativity of the phosphonic acid's tail. This insensitivity is likely due to a larger density of surface states and/or an unfavorable level alignment between the acid's HOMO and the GaO$_x$ surface states. Upon the grafting of phosphonic acids, PL measurements evidence a significant modification in the EQE of oxidized GaN$(1\bar{1}00)$ layers and nanowires. Besides the parasitic absorption caused by the organic SAM, these changes are attributed to the modification of the surface band bending and/or of surface states.

The large sensitivity of the GaN$(1\bar{1}00)$ electronic properties towards the adsorbate electronegativity is a key feature for the realization of chemo-sensors. In particular, selective sensing may be simply enabled by transforming the phosphonic acid's functional group into an anchor group that only binds to a targeted analyte \cite{Stine2010}. On the basis of these results, we expect a superior performance of the $(1\bar{1}00)$ surface with respect to the traditional $(0001)$ surface orientation \cite{Kang2005,Kang2007}. The use of the GaN$(1\bar{1}00)$ surface for sensing could be maximized using GaN nanowires which, in comparison to the established Si nanowire based biosensing technology \cite{He2010}, offer an improved biocompatibility \cite{Li2017} and near field light emission for optogenetic applications.  





\section{Acknowledgement}
We thank Katrin Morgenroth, Carsten Stemmler and Hans-Peter Sch\"onherr for their dedicated maintenance of the molecular beam epitaxy system and Sander Rauwerdink for his support with the sample treatments. We are indebted to Henning Riechert and Lutz Geelhaar for their continuous encouragement and support. We additionally thank Martin  Heilmann for the critical reading of the manuscript. Funding from the Bundesministerium für Bildung und Forschung through projects FKZ:13N13662, 13N13656, 13N13657 and, 13N13658 are gratefully acknowledged. Sergio Fernández-Garrido acknowledges the partial financial support received through the Spanish program Ramón y Cajal (co-financed by the European Social Fund) under grant RYC-2016-19509 from Ministerio de Ciencia, Innovación y Universidades.

\bibliography{GaN_BB_SAMs}

\end{document}


\title{External control of GaN band bending using phosphonate self-assembled monolayers --- Supplementary Information}

\author{T.~Auzelle} 
\affiliation{Paul-Drude-Institut für Festkörperelektronik, Leibniz-Institut im Forschungsverbund Berlin e.\,V., Hausvogteiplatz 5--7, 10117 Berlin, Germany}
\email{auzelle@pdi-berlin.de}

\author{F.~Ullrich}
\altaffiliation{Materials Science Department, Technische Universit\"at Darmstadt, Otto-Berndt-Strasse 3, 64287 Darmstadt, Germany}
\affiliation{InnovationLab, Speyerer Str. 4, 69115 Heidelberg, Germany}

\author{S.~Hietzschold}
\altaffiliation{Institute for High-Frequency Technology, Technische Universit\"at Braunschweig, Schleinitzstrasse 22, 38106 Braunschweig, Germany}
\affiliation{InnovationLab, Speyerer Str. 4, 69115 Heidelberg, Germany}

\author{C. Sinito}
\altaffiliation{Present address: Attolight AG, EPFL Innovation Park, Bât.\ D, 1015 Lausanne, Switzerland}
\affiliation{Paul-Drude-Institut für Festkörperelektronik, Leibniz-Institut im Forschungsverbund Berlin e.\,V., Hausvogteiplatz 5--7, 10117 Berlin, Germany}

\author{S.~Brackmann}
\affiliation{InnovationLab, Speyerer Str. 4, 69115 Heidelberg, Germany}

\author{W.~Kowalsky}
\altaffiliation{Institute for High-Frequency Technology, Technische Universit\"at Braunschweig, Schleinitzstrasse 22, 38106 Braunschweig, Germany}
\altaffiliation{Kirchhoff Institute for Physics, Heidelberg University, Im Neuenheimer Feld 227, 69120 Heidelberg, Germany}
\affiliation{InnovationLab, Speyerer Str. 4, 69115 Heidelberg, Germany}


\author{E.~Mankel}
\altaffiliation{Materials Science Department, Technische Universit\"at Darmstadt, Otto-Berndt-Strasse 3, 64287 Darmstadt, Germany}
\affiliation{InnovationLab, Speyerer Str. 4, 69115 Heidelberg, Germany}

\author{O. Brandt}
\affiliation{Paul-Drude-Institut für Festkörperelektronik, Leibniz-Institut im Forschungsverbund Berlin e.\,V., Hausvogteiplatz 5--7, 10117 Berlin, Germany}

\author{R.~Lovrincic}
\altaffiliation{Institute for High-Frequency Technology, Technische Universit\"at Braunschweig, Schleinitzstrasse 22, 38106 Braunschweig, Germany}
\altaffiliation{Present address: trinamiX GmbH, Industriestraße 35, 67063 Ludwigshafen, Germany}
\affiliation{InnovationLab, Speyerer Str. 4, 69115 Heidelberg, Germany}

\author{S.~Fern\'andez-Garrido}
\altaffiliation{Grupo de Electr\'onica y Semiconductores, Dpto. F\'isica Aplicada, Universidad Aut\'onoma de Madrid, C/ Francisco Tom\'as y Valiente 7, 28049 Madrid, Spain}
\email{sergio.fernandezg@uam.es}
\affiliation{Paul-Drude-Institut für Festkörperelektronik, Leibniz-Institut im Forschungsverbund Berlin e.\,V., Hausvogteiplatz 5--7, 10117 Berlin, Germany}

 \maketitle

\begin{figure}
\includegraphics[width = \linewidth]{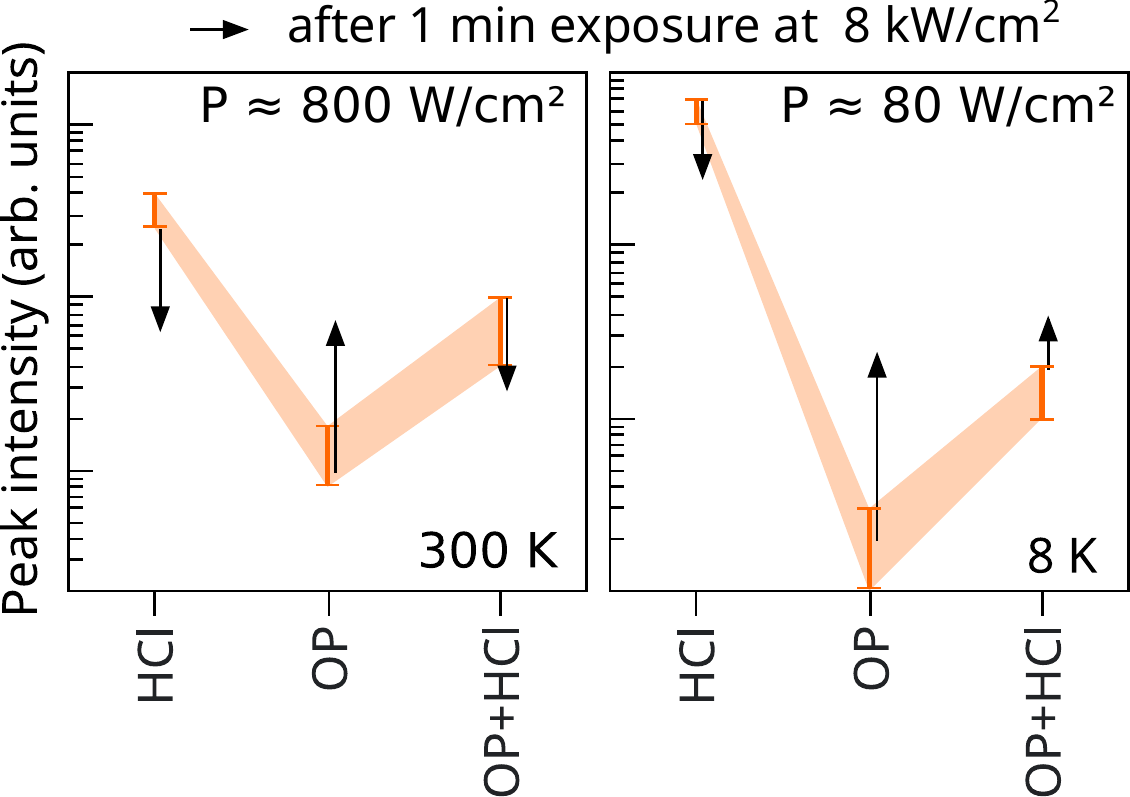}%
\caption{Peak PL intensity of the nanowire ensemble grown on Si after different surface treatments at 300 and 8\,K as indicated in the figures. The arrows indicate the intensity enhancement subsequent to a $1$\,min long laser exposure at $8$\,kWcm$^{-2}$. 
\label{fig:Plasma}}
\end{figure}

As described in the main text, the modulation of the nanowire photoluminescence intensity seen upon grafting of phosphonic acids points to the existence of a strong nonradiative decay taking place at the nanowire surface. This conclusion, however, differs from the results reported in Ref.\citenum{Zettler2016}, where the intense luminescence observed for ultrathin GaN nanowires reveals an extremely low recombination velocity associated to the surface states of air-oxidized GaN$(1\bar{1}00)$. The nanowires investigated here are, in contrast to Ref.\citenum{Zettler2016}, treated with an oxygen plasma (OP), where highly excited O radicals may create point defects in the vicinity of the surface. To test for this possibility, fresh pieces from the GaN nanowire ensemble grown on Si are measured either after an HCl dip, after OP exposure or, after OP exposure followed by an HCl dip. According to Ref.\citenum{Auzelle2019}, HCl etching provides flat band conditions for the GaN$(1\bar{1}00)$ surface, whereas the OP treatment gives an upward band bending. The related modification of the PL intensities at $8$ and $300$\,K are shown in Fig.\,\ref{fig:Plasma}. As can be observed, regardless of the temperature, the PL intensity clearly decreases after the OP treatment. The increased band bending caused by the OP treatment can partly explain the reduced PL intensity. However, an HCl dip performed after OP exposure does not fully recover the PL intensity. This result evidences the formation of point defects in the vicinity of the surface during the OP treatment. This modification of the surface concurs with the different effect of an intense laser exposure on the PL efficiency of the treated nanowires. As seen in Fig.\,\ref{fig:Plasma}, a laser irradiation at low temperature decreases the PL efficiency of HCl treated nanowires whereas it enhances the one of nanowires exposed to either OP or OP plus HCl. A further improvement of the grafting of phosphonic acids would thus require to get rid of the OP exposure step.

\bibliography{GaN_BB_SAMs}